\documentclass[usenatbib,onecolumn]{mn2e}
\RequirePackage[dvips]{graphicx}
\usepackage{amsmath,amssymb}
\usepackage{subfigure}

\newcommand{\nat}{Nature}
\newcommand{\apjl}{ApJ}
\newcommand{\apj}{ApJ}
\newcommand{\mnras}{MNRAS}
\newcommand{\aap}{A{\&}A}
\newcommand{\aapr}{A{\&}ARv}
\newcommand{\araa}{ARAA}

\title[Evolution of Thin Discs]{Evolution of Thin Discs with Partial Accretion}

\author[K. Y. Ek\c{s}i]{K. Y. Ek\c{s}i \\
Istanbul Technical University, Faculty of Arts and Letters, Physics Department, Maslak, 34469, Istanbul \\
E-mail: eksi@itu.edu.tr
}

\date{Released October 18, 2012}

\pagerange{\pageref{firstpage}--\pageref{lastpage}} \pubyear{2012}

\begin{document}

\maketitle

\label{firstpage}

\begin{abstract}
The viscous evolution of a thin disc around a central object is considered.
Such discs are described by self-similar solutions in which either all or none of the inflowing 
mass accretes.
An approximate solution for the partial accretion regime is constructed
by employing a prescription recently introduced for nonlinear heat conduction.
The solution is compared with numerical simulations demonstrating that the approximate solution describes the intermediate
asymptotic stage for partially accreting discs. 

\end{abstract}

\section{INTRODUCTION}

Accretion of gaseous material onto a central object 
is a process at the heart of many diverse astrophysical phenomena, 
from quasars to the formation 
of planetary systems \citep{fra02}. 
In many cases accretion is via a geometrically thin disc \citep{sha73,pri74b}.

The angular velocity of matter in the disc, $\Omega$, is keplerian ($\Omega_{\rm K}=\sqrt{GM_{\ast}/r^3}$ where $G$ is the gravitational constant, $M_{\ast}$ is the mass of the accreting object and $r$ is the cylindrical radial distance from the center of the star)
throughout the disc except in the narrow boundary layer 
where it matches to the angular velocity of the central object $\Omega_{\ast}$ at the inner radius of the disc, $r_{\rm in}$.

Accreting matter carries with it angular momentum  which is acquired 
by the central object if it is rotating slower than the 
keplerian angular velocity at the inner radius of the disc \citep{pri72}. 
The rate of material angular momentum flux by accretion is $\dot{L}_{\rm mat} =\dot{M} r^2 \Omega$ where $\dot{M}$ is the accretion rate. 
Viscous stresses also lead to a flux of angular momentum  $\dot{L}_{\rm vis} = 2 \pi r^3 \nu \Sigma d\Omega/dr$
where $\nu$ is the kinematic turbulent viscosity and $\Sigma$ is the surface mass density.
In the steady state the mass accretion rate, $\dot{M}$, and the total angular momentum flux, 
$\dot{L}=\dot{L}_{\rm mat} + \dot{L}_{\rm vis}$, in the disc are integration constants. For a keplerian disc the angular momentum balance 
leads to
\begin{equation}
\nu\Sigma = \frac{1}{3\pi} \left( \dot{M} -\frac{\dot{L}}{\sqrt{GM_{\ast}r}}  \right). 
\label{Jdot2}
\end{equation}
Usually one writes the total angular momentum flux in units of material stress for keplerian flow at the inner radius
\begin{equation}
\dot{L} \equiv \beta \dot{M} r_{\rm in}^2 \Omega_{\rm K}(r_{\rm in})
\label{beta}
\end{equation}
and the equation takes the more familiar form
\begin{equation}
\nu\Sigma = \frac{\dot{M}}{3\pi} \left( 1 - \beta \sqrt{\frac{r_{\rm in}}{r}}  \right). 
\label{Jdot3}
\end{equation}

Within the disc the viscous  and the material torques can have different ratios 
though the total angular momentum flux is a constant determined 
by the fastness parameter $\omega_{\ast} \equiv \Omega_{\ast}/\Omega_{\rm K}(r_{\rm in})$. 
If the central object is rotating slowly,
the angular velocity reaches a maximum value before matching the angular velocity of the star.  
At this point the viscous torque, 
as it depends on the radial angular velocity gradient, vanishes 
and the material torque
is equal to the total angular momentum flux. If the boundary layer is narrow this allows one to estimate  
 $\beta \simeq 1$ for slowly rotating objects.

If the disc is 
truncated beyond the stellar radius e.g.\ by the strong magnetic dipole field of the star, it is possible 
without stellar break-up,
that the angular velocity of the star is much greater 
than the keplerian 
angular velocity at the inner radius of the disc  ($\omega_{\ast}>1$). In such cases the mass flux
can diminish due to the centrifugal barrier \citep{ill75} and the viscous 
torque is negative i.e.\ the angular momentum is transferred from the star to the disc ($\beta<0$). 
\citet{sun77} gave a steady state solution
for such ``dead discs'' in which mass flux (and so the material torque) is zero.

The steady state solutions of \citet{sha73} and \citet{sun77} have time-dependent counterparts as given by \citet{pri74a}.
These are self-similar solutions and extend to the origin, $r=0$.
In the first solution the mass accretion rate declines with a power-law in time while the angular momentum of the disc 
is constant ($\dot{L}=0$) \citep{can90}. In the second solution the mass of the disc is constant ($\dot{M}=0$) but angular momentum of the disc increases with the viscous torque at the inner boundary \citep{pri91}. These extreme boundary conditions and the extension of the disc to $r=0$ in the solution
manifest the necessity that no length scale should exist in order that a self-similar solution can be found.

In reality, as the disc is not infinitely thin, matter vertically away from the disc plane can accrete 
even when matter is propelled at the midplane. Thus there should be a transition stage between these 
full accretion and full propeller stages during which the central object will accrete a fraction of 
the inflowing material \citep{men99,eks10,rom04,ust06} depending on 
the fastness parameter. 
Furthermore, it is expected that an episodic accretion stage \citep{spr93,dang11,dang12} 
will be realized near the transition stage separating the full accretion and full propeller stages. 
Such a stage can not be described as a succession of steady states as $\dot{M}$ is an integration constant and
hence whatever the accretion rate in the disc should be the rate of accretion onto the star.
The self-similar solutions because they correspond to extreme boundary conditions, 
either $\dot{L}=0$ or $\dot{M}=0$, also can not accommodate the continuum of angular 
momentum flux per mass accretion rate, $\dot{L}/\dot{M}$, to be realized in the continuous transition between 
the full propeller and full accretor stages.

In this work we construct an approximate self-similar time dependent solution via a prescription 
recently used by \citet{eks09} for self-similar evolution of temperature distribution
in a semi-infinite rod with partial heat insulation at one boundary. In the next section disc structure equations and the self-similar 
solutions of \citet{pri74a} is reviewed. In \S 3 the ``nonlinear superposition'' method of \citet{eks09} 
is applied to these solutions to
construct a self-similar approximate solution with arbitrary values of $\beta$. In \S 4  we demonstrate
the accuracy of the approximate analytical solution by comparing it with numerical results.
In the final section we discuss some implications for neutron star soft X-ray transients.

\section{EVOLUTION OF THE DISC}
We consider the evolution of a non-self-gravitating viscous thin disc in the gravitational field
of a central mass $M_{\ast}$. 

\subsection{Thin Disc Equations}

The conservation of mass in cylindrical coordinates
reads%
\begin{equation}
\frac{\partial \Sigma }{\partial t}=\frac{1}{2\pi r}\frac{\partial \dot{M}}{%
\partial r}  \label{cont1}
\end{equation}%
where $\Sigma $ is the surface mass density and%
\begin{equation}
\dot{M}=2\pi r\Sigma \left( -v_{r}\right)  \label{int1}
\end{equation}%
is the mass flow rate, $v_{r}$ being the radial velocity. The conservation
of angular momentum reads%
\begin{equation}
\frac{\partial }{\partial t}(r^{2}\Omega \Sigma )=\frac{1}{2\pi r}\frac{%
\partial \dot{L}}{\partial r}.  \label{ang1}
\end{equation}%
where $\Omega $ is the angular velocity of the fluid and 
\begin{equation}
\dot{L}=\dot{M}r^{2}\Omega +2\pi r^{3}\nu \Sigma \frac{\partial \Omega }{\partial r}   
\label{int2}
\end{equation}%
is the angular momentum flux (torque). The first term is the material torque 
and the second term is the viscous torque 
where $\nu $ is the turbulent kinematic viscosity.

The form of the conservation equations (\ref{cont1}) and (\ref{ang1}) make
it clear that in the steady state the accretion rate $\dot{M}$ and angular
momentum flux $\dot{L}$ are two integrals of motion. Assuming the matter in the disc is in Keplerian orbits, $\Omega=\Omega_{\rm K}$ where $\Omega_{\rm K} \equiv \sqrt{GM_{\ast}/r^3}$, Equation (\ref{ang1}) with the help of Equation~(\ref{cont1}) 
reduces to
\begin{equation}
\dot{M}=6\pi r^{1/2}\frac{\partial }{\partial r}\left( \nu \Sigma r^{1/2}\right)
\label{acc_rate}
\end{equation}
This then can be plugged into Equation~(\ref{cont1}) to give the diffusion equation 
\begin{equation}
\frac{\partial \Sigma }{\partial t}=\frac{3}{r}\frac{\partial }{\partial r}%
\left[ r^{1/2}\frac{\partial }{\partial r}\left( \nu \Sigma r^{1/2}\right) %
\right] 
\label{diff1}
\end{equation}
for the evolution of the surface mass density.
The equation is linear if $\nu$ does not depend on $\Sigma$.
In general the viscosity depends on $\Sigma$ and the equation is nonlinear. 

The rest of the disc structure equations  are as follows \citep{fra02}: We employ \citet{sha73} turbulent viscosity prescription $\nu=\alpha c_{\rm s} H$ where $\alpha \sim 0.01-0.1$ is the dimensionless viscosity parameter, $c_{\rm s}=\sqrt{P/\rho}$ is the sound speed, $H=c_{\rm s}/\Omega_{\rm K}$ is the disc scale height and $P$ is the pressure. 
The density is then given by $\rho=\Sigma/H$. Assuming the heat produced by viscous processes are radiated locally, the energy balance can be written as $4\sigma_{\rm SB} T^4/3\Sigma \kappa= \frac98 \nu \Sigma \Omega_{\rm K}^2$  where $T$ is the temperature at the disc mid plane and $\kappa$ is the Rosseland mean opacity. These equations are then supported by the equation of state $P=P(\rho,T)$ and a prescription for the opacity $\kappa = \kappa(\rho,T)$. The pressure has contributions from both the gas pressure, $P_{\rm gas}=\rho k_{\rm B} T/\bar{\mu} m_{\rm p}$ where $\bar{\mu}$ is the mean molecular weight, and radiation pressure $P_{\rm rad}=4 \sigma_{\rm SB} T^4/3c$. We assume the gas pressure dominates the radiation pressure throughout the disc. Opacity is in general of the form $\kappa=\kappa_0 \rho^a T^b$ where $\kappa_0$, $a$ and $b$ are constants depending on the dominating opacity regime (e.g.\ $a=b=0$ and $\kappa_0=0.4$ cm$^2$ g$^{-1}$ for electron scattering). These algebraic disc structure equations can be solved among themselves first to obtain viscosity (and other variables) in terms of $r$ and $\Sigma$ in the form 
\begin{equation}
\nu = C r^p \Sigma^q
\label{viscosity}
\end{equation}
where $C$, $p$ and $q$ are constants determined by the dominant opacity regime and pressure (see e.g.\ \citet{can90,ert09}). This then can be plugged in 
the diffusion equation (\ref{diff1}) to obtain an equation containing $\Sigma$ only. For discs in which gas pressure and electron scattering opacity dominates $p=1$ and $q=2/3$ \citep{can90}. In comparing the analytical solutions with the numerical solutions we assume this regime to prevail throughout the disc though it is well known that different opacity and pressure regimes prevail at different locations of the disc depending on temperature $T(r,t)$. 

\subsection{Pringle Solutions}

In order to apply self-similarity methods we render the equation dimensionless  via
$R=r/r_0$, $\tau=t/t_0$ and $\sigma=\Sigma/\Sigma_0$, and define $\nu_0=Cr_0^p \Sigma_0^q$. Choosing $t_0=4r_0^2/3\nu_0$ we obtain
\begin{equation}
\frac{\partial \sigma }{\partial \tau }=\frac{4}{R}\frac{\partial }{\partial
R}\left[ R^{1/2}\frac{\partial }{\partial R}\left( R^{p+1/2} \sigma^{q+1}\right) \right]. 
\label{diff2}
\end{equation}
Self-similarity methods \citep{zel67} provide two solutions of this equation as first found by 
\citet{pri74a} (see also \citet{pri74b,tan11,min91,min93}). The first solution which we call the ``full accretor solution'', for reasons to be explained in the following subsection, is
\begin{equation}
\sigma(R,\tau)=
\begin{cases}
(1+\tau)^{-\frac{5}{5q-2p+4}}\left( R(1+\tau)^{-\frac{2}{5q-2p+4}}\right)
^{-\frac{p}{q+1}}\left[ 1-k_{p,q}\left( R(1+\tau)^{-\frac{2}{5q-2p+4}}\right) ^{2-%
\frac{p}{q+1}}\right]^{1/q}, & \mbox{if }q\neq 0,   \\ 
(1+\tau)^{\frac{5-2p}{2(p-2)}}R^{-p}\exp \left( -%
\frac{R^{2-p}}{4\left( p-2\right) ^{2}(1+\tau)}\right), & \mbox{if }q=0%
\end{cases}
\label{sol1}
\end{equation}
where 
\begin{equation}
k_{p,q}=\frac{q}{( 4q-2p+4) ( 5q-2p+4)}.
\end{equation}
The second solution which we call the ``full propeller solution'' is
\begin{equation}
\sigma(R,\tau) =
\begin{cases}
(1+\tau)^{-\frac{4}{4q-2p+4}}\left( R(1+\tau)^{-\frac{2}{4q-2p+4}}\right)
^{-\frac{p+1/2}{q+1}}\left[ 1-k_{p,q}\left( R(1+\tau)^{-\frac{2}{4q-2p+4}}\right) ^{%
\frac{5}{2}-\frac{p+1/2}{q+1}}\right] ^{1/q}, & \mbox{if }q\neq 0,   \\ 
(1+\tau)^{\frac{3-2p}{2(p-2)}}R^{-p-\frac{1}{2}%
}\exp \left( -\frac{R^{2-p}}{4 \left( p-2\right)^{2}(1+\tau)}\right) 
, & \mbox{if }q=0.
\end{cases}
\label{sol2}
\end{equation}
Note that Equation~(\ref{diff1}) is
symmetric under translations in time and hence the solutions can also be shifted in time.
We have exploited this to write the original solutions shifted as $\tau \rightarrow \tau + 1$.
Note also that the solutions for the linear cases ($q=0$) can be found from the nonlinear cases by referring 
$\lim_{q\rightarrow 0}(1+Aq)^{1/q}=\exp{A}$. In the linear solutions the $p=2$ case should be handled separately as is done by \citet{pri74a}. We do not write the solution for this case as it would not contribute to the main motivation of the paper. 

In the non-dimensionalization process we have employed four quantities, $\Sigma_0$, $r_0$, $t_0$ and $\nu_0$, related by two equations $\nu_0=Cr_0^p\Sigma_0^q$ and $t_0=4r_0^2/3\nu_0$. This means we are free to attribute any value for two of the variables e.g.\ $\Sigma_0$ and $r_0$. We fix these two quantities in terms of the initial mass and angular momentum of the disc  \citep[see][]{ert09}.

\subsection{Properties of solutions}

The Pringle solutions given in Equations (\ref{sol1}) and (\ref{sol2}) describe the evolution of thin discs with free outer boundaries and thus are more appropriate for discs
that are not truncated e.g.\ by tidal torques. 
The physical meaning of the first solution (full accretor)  was clarified by \citet{can90} who modeled the evolution of a disc formed by tidally disrupted star around a supermassive black hole. Soon afterwards \citet{pri91} used the second solution for describing the evolution of circumbinary discs. \citet{min91} used the first solution for the evolution of dwarf nova discs. \citet{min93} used it for the evolution of the putative disc around the neutron star assumed to form in SN 87A.
Since the original work of \citet{cha00} the first solution has been employed for describing fallback discs 
\citep[e.g.][]{per00,ert09} around young neutron stars. \citet{eks03} employed both two solutions, the first one for the accretion and the second one for the propeller stages of the system, respectively.

The properties of the solutions given in Equations (\ref{sol1}) and (\ref{sol2}) are  
understood and classified in terms of the time evolution they result for 
the mass
\begin{equation}
M_{\mathrm{d}}=\int_0^{r_{\rm out}} \Sigma \cdot 2\pi rdr
\label{M_d}
\end{equation}
and the angular momentum
\begin{equation}
L_{\mathrm{d}}=\int_0^{r_{\rm out}} r^{2}\Omega_{\mathrm K} \Sigma \cdot 2\pi rdr,
\label{J_d}
\end{equation}
of the disc. Here $r_{\rm out}$ is the freely expanding outer boundary of the disc defined as the location where
the square bracket terms in the solutions given in Equations (\ref{sol1}) and (\ref{sol2}) vanish. 
The lower limit of the integrals is zero as the solutions extend to the origin. 
A real disc will have a finite inner radius and so the integrals then can provide the correct parameters as long as this inner radius is much smaller than the outer radius.

\begin{figure*}
\includegraphics[width=0.45\textwidth]{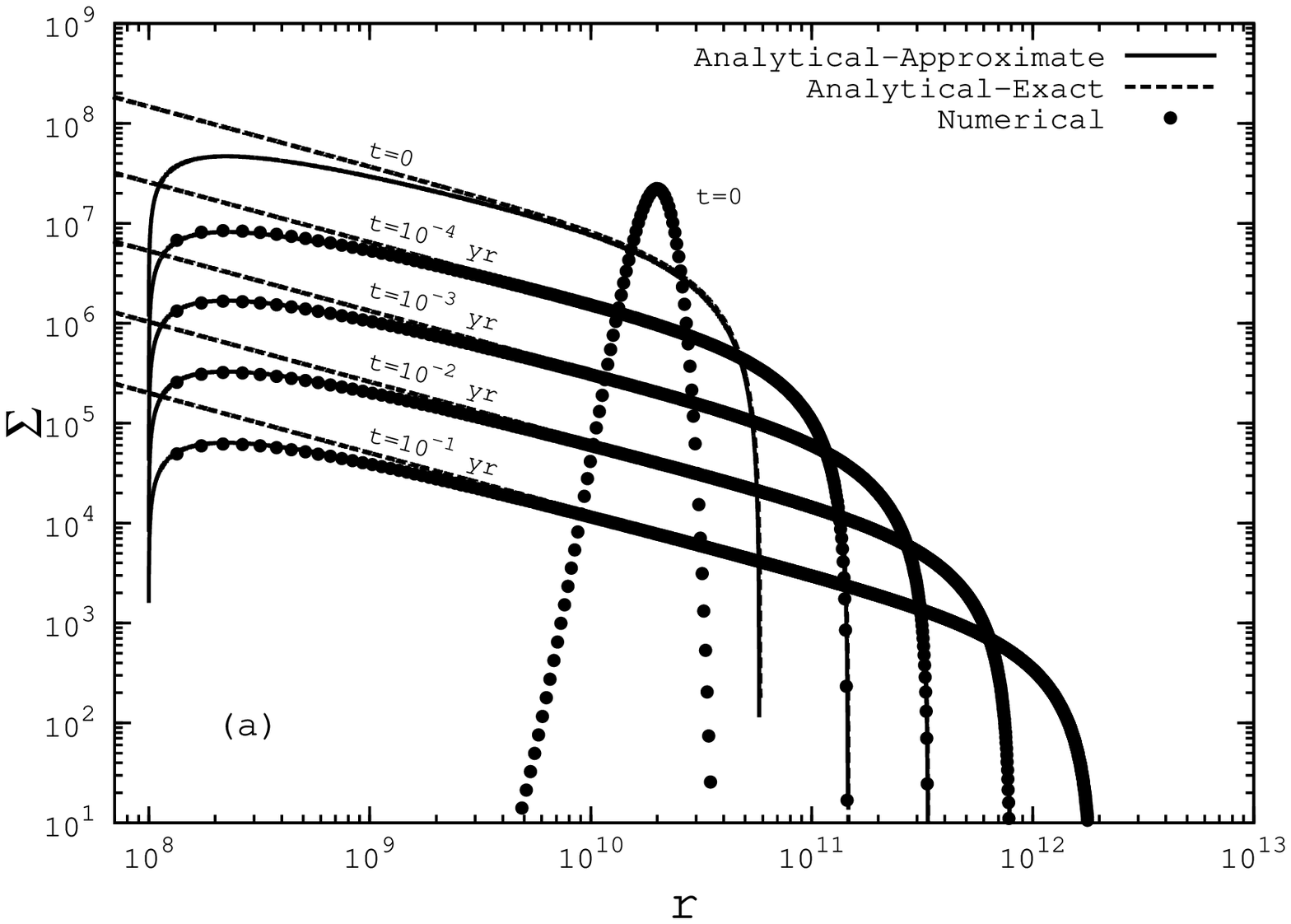}
\includegraphics[width=0.45\textwidth]{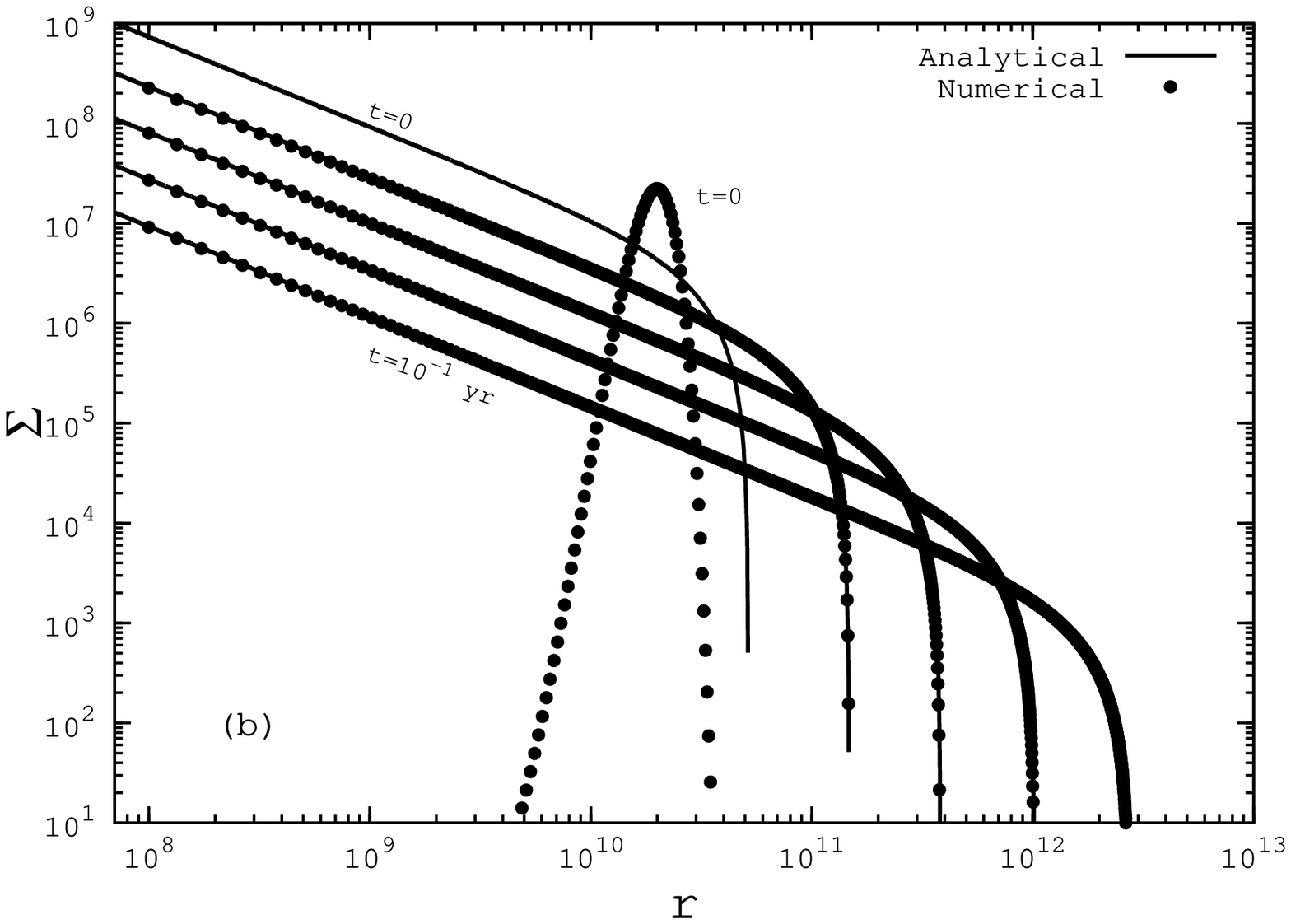}
\caption{Comparison of the analytical and numerical solutions of the diffusion equation~(\ref{diff1}) for the full accretor (panel a) and full propeller (panel b) inner boundary conditions, $\Sigma=0$ and $\partial (r^{1/2}\nu \Sigma)/\partial r=0$ at $r=r_{\rm in}$, respectively. The filled circles show the numerical solution which starts from a Gaussian surface density distribution. On the left panel the solid line shows the approximate analytical solution given in Equation~(\ref{approximate}) for $n=1$. The dashed line shows the exact Pringle solution given in Equation~(\ref{solut1}).}
\label{disc_uni}
\end{figure*}

\subsubsection{Full Accretor Solution}

According to the first solution given in Equation~(\ref{sol1}), 
the outer boundary of the disc evolves as $r_{\rm out}\propto (1+\tau)^{2/(5q-2p+4)}$. The mass of the disc
decays with $M_{\mathrm d} \propto (1+\tau)^{-1/(5q-2p+4)}$ so that mass flux out of the disc is
\begin{equation}
\dot{M}_{\mathrm d}=\dot{M}_0 (1+\tau)^{-1-\frac{1}{5q-2p+4}},
\label{Mdot_d}
\end{equation}
where $\dot{M}_0\equiv 3\pi \nu_0 \Sigma_0$ is the initial mass flow rate in the disc. In this solution the angular momentum of the disc
is constant ($\dot{L}_{\mathrm{d}}=0$): the outer radius of the disc expands at the exact rate to take up the angular momentum  lost from the inner disc by the accreting mass.

\citet{can90} showed that this solution describes the evolution of the disc with the absorbing inner boundary condition $\Sigma (r_{\mathrm{in}},t)=0$ at the inner radius of the disc, $r_{\mathrm{in}}$. In this solution $\sigma \propto R^{-p/(q+1)}$ for most part of the disc except near the outer boundary which the terms in square brackets handle. As shown in the next section this also is the radial profile in the steady-state counterpart \citep{sha73} away from the inner boundary of the disc. 
The self-similar solutions are thus said to describe a quasi-static regime i.e.\ successive steady-states with decreasing accretion rates \citep{pri74a,cha05}.
According to this solution the disc extends to the origin and $\sigma$ blows out there. In the numerical solution $\sigma$ overturns trying to match the inner boundary condition. 
As first remarked by \citet{ert09} the time shifted solutions where $r_0$ and $\Sigma_0$ are fixed by referring to the initial mass and angular momentum in the disc gives a better match to the numerical solution. The numerical solution settles to the self-similar solution in a few diffusive timescales. As such self-similar solutions describe the intermediate asymptotic regime \citep{bar96} as they prevail after the initial conditions are forgotten and before the disc vanishes completely.

\subsubsection{Full Propeller Solution}

According to second solution given in Equation (\ref{sol2}) but shifted in time the outer radius evolves as $r_{\rm out}\propto (1+\tau)^{2/(4q-2p+4)}$. The mass of the disc is constant ($\dot{M}_{\mathrm{d}}=0$)
and the angular momentum of the disc increases as $L_{\mathrm d} \propto (1+\tau)^{1/(4q-2p+4)}$ 
so that the angular momentum flux evolves as 
\begin{equation}
\dot{L}_{\mathrm d}=\dot{L}_0(1+\tau)^{\frac{1}{4q-2p+4}-1}.
\end{equation}
where $\dot{L}_0=\dot{M}_0 \sqrt{GMr_0}$ is the initial angular momentum flux. \citet{pri91} showed that this describes the intermediate asymptotic regime of the evolution of a disc with the boundary condition $\partial(\nu\Sigma r^{1/2})/\partial r=0$ at $r=r_{\mathrm{in}}$ which allows no mass flux out of the disc from the inner boundary (see Eqn.(\ref{acc_rate})). \citet{pri81} mentioned that this solution could describe the propeller regime \citep{ill75} of a disc in which the centrifugal barrier set by a rapidly rotating magnetosphere prohibits accretion. In this solution radial profile is $\sigma \propto R^{-(p+1/2)/(q+1)}$ which, again, extends to the origin and blows up. This also is the radial profile in the steady-state counterpart \citep{sun77} as shown in the next section. 
The torque given in Equation~(\ref{int2}) consists of two parts, a material and a viscous torque. As there is no matter flux out of the disc, in this solution, the angular momentum flux above is due to the viscous torque only. In order to take up the angular momentum transfered to the disc by this torque, the outer radius of the disc expands more rapidly in this solution than it does in the full accretor solution (see Figure~\ref{disc_uni}).

\section{AN APPROXIMATE SOLUTION WITH GENERAL BOUNDARY CONDITIONS}

The Pringle solutions correspond to two extreme stages: either \emph{all} matter reaching the 
inner radius of the disc accretes while the angular momentum of the disc is constant 
(first solution) or \emph{none} accretes (mass of the disc is constant) while the angular 
momentum of the disc increases with the viscous torque from the inner boundary (second solution). 

It would be desirable to have a solution which could accommodate the continuous range of $\dot{L}/\dot{M}$.
In the linear case ($q=0$) such a solution can be constructed by superposing the solutions; 
namely one can multiply the solutions with appropriate constants and adds them up. In the nonlinear case superposition
can not be applied. 

The heat equation $\partial_t u=\partial_{x} \chi \partial_{x} u$ with power-law heat conduction coefficient ($\chi=\chi_0 u^q$) is mathematically a special case (obtained for $2p=3q$) of the diffusion equation
describing the evolution of the disc. 
There are self-similar solutions describing the evolution of temperature on a semi-infinite rod $x=[0,\infty)$ \citep{zel67} which are then special cases of the Pringle solutions.
The first solution is a special case of the full accretion solution and accommodates $u=0$ at $x=0$
and so all heat inflowing there is absorbed.
The other solution is a special case of the full propeller solution and accommodates $\partial u/\partial x=0$ at $x=0$ which allows no heat flux through the inner boundary (perfect insulator).  Recently, \citep{eks09} constructed an approximate self-similar solution for the partial insulator boundary condition at $x=0$.
In this section we are going to employ the same prescription to the Pringle solutions to obtain an approximate solution for the partial accretion case.

\subsection{Solutions in terms of fluxes}

As a first step, we write the first and second solutions in terms of mass and angular momentum flux, respectively. 
It is possible to write the first solution given in 
Equation (\ref{sol1}) in terms of the dimensionless mass flux
$\dot{m} \equiv \dot{M}/\dot{M}_0=(1+\tau)^{-1-\frac{1}{5q-2p+4}}$ as
\begin{equation}
\sigma =
\begin{cases}
\left( R^{-p}\dot{m}\right) ^{\frac{1}{q+1}}\left[ 1-k_{p,q}\frac{R^{2-p}}{1+\tau}
\left( R^{-p}\dot{m}\right)^{\frac{1}{q+1}-1}\right] ^{1/q} , & \mbox{if }q\neq 0,   \\ 
\left( R^{-p}\dot{m}\right)\exp \left( -\frac{R^{2-p}}{4\left( p-2\right) ^{2}(1+\tau)}\right), & \mbox{if }q=0%
\end{cases}
\label{solut1}
\end{equation}
The second solution given in Equation (\ref{sol2}), in terms of dimensionless angular momentum flux $\dot{\ell}\equiv \dot{L}/\dot{L}_0=(1+\tau )^{\frac{1}{4q-2p+4}-1}$, can be written as
\begin{equation}
\sigma =
\begin{cases}
\left[ R^{-p}(-\dot{\ell}R^{-1/2})\right] ^{\frac{1}{q+1}}\left[ 1-k_{p,q}\frac{R^{2-p}}{1+\tau}
\left( R^{-p}(-\dot{\ell}R^{-1/2})\right)^{\frac{1}{q+1}-1}\right] ^{1/q}, & \mbox{if }q\neq 0,   \\
\left[ R^{-p}(-\dot{\ell}R^{-1/2})\right] \exp \left( -\frac{R^{2-p}}{4(p-2)^{2}(1+\tau) }\right) 
, & \mbox{if }q=0
\end{cases}
\label{solut2}
\end{equation}
Note that the viscous torque is negative when this solution is valid so that $-\dot{\ell}$ is actually positive.

We have eliminated \emph{most} occurrences of $\tau$ in favor of mass flux in the full accretor solution and angular momentum flux in the full propeller solution. Note the ``dual form'' of the solutions ($\dot{m}\rightarrow -\dot{\ell}R^{-1/2}$) which would not exist if we insisted in eliminating \emph{all} occurrences of $\tau$ in favor of $\dot{m}$ and $\dot{\ell}$, respectively. This leads to another useful property: 
\emph{The solutions as they are given in Equations (\ref{solut1}) and (\ref{solut2}) satisfy the diffusion equation (\ref{diff2}) even when  $\dot{m}$ and $\dot{\ell}$ are defined as
\begin{equation}
\dot{m} = \dot{m}_0 (1+\tau)^{-\mu}, \qquad \mu = 1+\frac{1}{5q-2p+4}
\label{mdot}
\end{equation}
and
\begin{equation}
\dot{\ell} = \dot{\ell}_0 (1+\tau)^{-\lambda}, \qquad \lambda=1-\frac{1}{4q-2p+4}
\label{ldot}
\end{equation}
i.e.\ when they are multiplied by some arbitrary constants $\dot{m}_0$ and $\dot{\ell}_0$. This, in the linear cases ($q=0$), is equivalent to multiplying the solution itself with a constant as the flux is then a multiplier of the solution.}
This property, indispensable for the superposition procedure, would not be valid if we insisted in eliminating all occurrences of $\tau$ in favor of the fluxes.

\subsection{Comparing with the steady state solution}

The steady-state solution of $\Sigma$ can be found from  Equations (\ref{Jdot2}) and (\ref{viscosity}). In dimensionless form this reads
\begin{equation}
\sigma =\left[ R^{-p}\left( \dot{m}-\dot{\ell}R^{-1/2}\right) \right] ^{\frac{1}{q+1}}.
\label{steady}
\end{equation}
Here $\dot{m}$ and $\dot{\ell}$ are integration constants.
The solutions in terms of fluxes, given in equations (\ref{solut1}) and (\ref{solut2}), are equivalent to the original solutions given in Equations (\ref{sol1}) and (\ref{sol2}), respectively. The meaning of the self-similar solutions when written in the latter form is
much transparent showing clearly that the self-similar solutions are special solutions in the sense that they each have a single integration constant (in space) whereas any ratio between $\dot{m}$ and $\dot{\ell}$ is possible in the the steady state solution given in Equation (\ref{steady}). The steady state counterparts of the solutions given in Equations (\ref{solut1}) and (\ref{solut2}) would be 
$\sigma =[ R^{-p}( \dot{m})]^{1/(q+1)}$ and $\sigma =[R^{-p}( -\dot{\ell}R^{-1/2})]^{1/(q+1)}$, respectively. 
The first of these is the full accretor solution with no viscous torque and inner radius at the origin. 
The other limiting case, $\sigma =[R^{-p}( -\dot{\ell}R^{-1/2})]^{1/(q+1)}$, obtained for  $\dot{m}=0$ from Equation~(\ref{steady}), is known as the ``dead disc'' solution of \citet{sun77}. Such a disc is expected to form in the propeller regime \citep{ill75} though it would not be steady.

\subsection{Nonlinear ``superposition''} 

The idea of going back from the limiting cases, to the general solution given in Equation~(\ref{steady}) suggests a ``superposition'' principle
\begin{equation}
\left. 
\begin{array}{lll}
\sigma &=& \left[R^{-p}\left(\dot{m} \phantom{-\dot{\ell}R^{-1/2}}\right)\right]^{1/(q+1)} \\ 
\sigma &=& \left[R^{-p}\left(\phantom{\dot{m}}-\dot{\ell}R^{-1/2}\right)\right]^{1/(q+1)}
\end{array}%
\right\} \Longrightarrow \sigma =\left[ R^{-p}\left(\dot{m}-\dot{\ell}R^{-1/2}\right)
\right]^{1/(q+1)}
\end{equation}
which reduces to ordinary summation in the linearity limit ($q=0$).
Using this procedure, the solutions given in Equations (\ref{solut1}) and (\ref{solut2}) can be ``superposed'' as 
\begin{equation}
\sigma =
\begin{cases}
\left[ R^{-p}(\dot{m}-\dot{\ell}R^{-1/2})\right] ^{\frac{1}{q+1}}\left[ 1-k_{p,q}\frac{R^{2-p}}{1+\tau}
\left( R^{-p}(\dot{m}-\dot{\ell}R^{-1/2})\right)^{\frac{1}{q+1}-1}\right] ^{1/q}, & \mbox{if }q\neq 0,   \\
\left[ R^{-p}(\dot{m}-\dot{\ell}R^{-1/2})\right] \exp \left( -\frac{R^{2-p}}{4(p-2)^{2}(1+\tau) }\right) 
, & \mbox{if }q=0.
\end{cases}
\label{unified}
\end{equation}
Clearly, $\dot{\ell}_0=0$ case will give the ``full accretor'' and $\dot{m}_0=0$ will give the ``full propeller'' solution given in Equations (\ref{solut1}) and (\ref{solut2}), respectively.

If we plug in the expression given in Equation (\ref{unified}) into the diffusion equation (\ref{diff2}) we find that
the condition for satisfying it reduces to
\begin{equation}
\frac{1}{q+1}(1+\tau)^{-1}f(R,\tau)\left[\left(\frac{\ddot{m}(1+\tau)}{\dot{m}} + \mu \right)R^2 \dot{m}^2+\left(\frac{\ddot{m} (1+\tau)}{\dot{m}} + \frac{\ddot{\ell}(1+\tau)}{\dot{\ell}}+2 \right)R\dot{m}\dot{\ell}+ \left(\frac{\ddot{\ell}(1+\tau)}{\dot{\ell}} +\lambda \right) \dot{\ell}^2 \right]=0.
\label{cond1}
\end{equation} 
where $f(R,\tau)$ is some complicated function not very important for our discussion.
The first and the third terms in square brackets vanish upon using Equations~(\ref{mdot}) and (\ref{ldot}), respectively. The second term does not vanish and the condition for Equation (\ref{unified}) to satisfy the diffusion equation (\ref{diff2}) reduces to
\begin{equation}
\frac{k_{p,q}}{q+1}\frac{\dot{m}\dot{\ell}}{1+\tau}R f(R,\tau)=0
\label{cond2}
\end{equation} 
where we have used $2-\mu-\lambda \equiv k_{p,q}$. 
Thus, the expression given in Equation (\ref{unified}) is not an exact solution unless $\dot{m}_0=0$ or $\dot{\ell}_0=0$ 
(reducing to one of the already known solutions given in Equations (\ref{solut1}) and (\ref{solut2}), respectively), or when $k_{p,q}=0$ which is accomplished when $q=0$ i.e.\ the linearity limit. It thus appears that we have not gained much with the unified expression  given in Equation~(\ref{unified}) as it
is an exact solution only in the already known cases. 
Yet, we will show in the following that the unified expression given in Equation (\ref{unified}) not only illuminates the meaning and complementarity of the known solutions but also motivates a very accurate approximate solution describing the partial accretion regime becoming exact for $\tau\rightarrow \infty$, practically achieved in a few viscous time-scales. 

\subsection{Why not an exact solution?}

Could there be an exact solution of the diffusion equation which could accommodate boundary conditions in which 
angular momentum flux per unit mass accretion rate could take any value?
In order that a self-similar solution is found no scale should exist in the problem \citep{bar96}. 
The expression given in Equation (\ref{unified}) or any expression with both mass accretion rate 
and angular momentum flux having finite values 
can not be an exact solution because of the presence of the length scale 
$l=\Sigma(0)/\Sigma'(0)\propto (\dot{L}/\dot{M})^2/GM$ creeping into the problem 
through the boundary condition and destroying the very condition for the existence of 
a self-similar solution.  Accordingly, the exact solutions given in Equations (\ref{solut1}) 
and (\ref{solut2}) are found either when $l=0$ ($\dot{L}=0$) or 
when $l\rightarrow \infty$ ($\dot{M}=0$). Any attempt to find a  
solution with a finite inner radius also would suffer from the same problem as this also would introduce 
a length scale into the problem \citep{spr01}  except for the case $2p=3q$ where the equation is symmetric under translations so that the origin can be shifted. We stress that a solution obtained by the 
non-linear superposition of self-similar solutions is not equivalent to shifting 
the origin of the coordinates.

The superposition procedure in the linear case also introduces a length scale into the problem. 
Yet the superposed expression ($q=0$ case in Equation~(\ref{unified})) \emph{is} an exact solution.
The presence of a scale  does not hinder the exact solution in the linear case
because, as hinted in Equation (\ref{cond2}), the length scale is coupled to 
the ``non-linearity parameter'' $q$: vanishing of either renders the expression 
given in Equation~(\ref{unified}) exact.

\subsection{A more useful ``solution''}

Note that $\dot{m}$ and $\dot{\ell}$ have different time dependencies, as given by Equations (\ref{mdot}) and (\ref{ldot}), respectively. This implies the inner radius of the disc  in the ``solution'' given in Equation (\ref{unified}) is time-dependent. Note, however, that the generalization to the expression given in Equation (\ref{unified}) is attained from solutions with zero inner radius. 
For example, the solution with no viscous torque but finite inner radius would be
\begin{equation}
\sigma =\left[ R^{-p}\dot{m}\left(1 -(R_{\rm in}/R)^{1/2}\right) \right] ^{\frac{1}{q+1}}
\label{steady2}
\end{equation}
where $R_{\rm in}=r_{\rm in}/r_0$ \citep{pri81}, rather than $\sigma =[R^{-p}\dot{m}]^{\frac{1}{q+1}}$. 
In order to accommodate cases with finite angular momentum flux 
we refer to Equation (\ref{beta}) defining the dimensionless torque $\beta$. This in dimensionless form can be written as
\begin{equation}
\dot{\ell}\equiv \beta \dot{m} \sqrt{R_{\rm in}}
\label{n}
\end{equation}
(recall Eqn.~(\ref{beta})), and 
write the ``solution'' given in Equation (\ref{unified}) in the form
\begin{equation}
\sigma =
\begin{cases}
\left[ R^{-p}\dot{m}(1-\beta(R_{\rm in}/R)^{1/2})\right] ^{\frac{1}{q+1}}\left[ 1-k_{p,q}\frac{R^{2-p}}{1+\tau}
\left( R^{-p}\dot{m}(1-\beta(R_{\rm in}/R)^{1/2})\right)^{\frac{1}{q+1}-1}\right] ^{1/q}, & \mbox{if }q\neq 0,   \\
\left[ R^{-p}\dot{m}(1-\beta(R_{\rm in}/R)^{1/2})\right] \exp \left( -\frac{R^{2-p}}{4(p-2)^{2}(1+\tau)}\right) 
, & \mbox{if }q=0
\end{cases}
\label{approximate}
\end{equation}
which accommodates the no-viscous-torque inner boundary condition for $\beta=1$.
The left panel of Figure~\ref{disc_uni} shows this \emph{approximate} solution 
and the exact Pringle solution given in Equation~(\ref{solut1}) together with the numerical solution obtained with the boundary condition $\nu \Sigma (r_{\mathrm{in}},t)=0$. 

The numerical solution (see the next section for the details of the numerical method) is 
started from a Gaussian initial profile. In comparing the solutions 
we have calculated the mass and angular momentum in this Gaussian surface density distribution 
and used these values for the analytical solution. No fitting procedure is applied. 
As first shown by \citet{can90} and \citet{pri91} the figures demonstrate that once 
the initial conditions are forgotten in a 
few viscous timescales, the evolution of the disc proceeds self-similarly. 
It is instructive to compare the exact Pringle solution given in Equation~(\ref{solut1}) 
and the approximate solution given in Equation~(\ref{approximate}): The approximate solution 
describes the inner region of the disc much better though they are both satisfactory for 
the intermediate and outer regions. The mass flow rate in the numerical and analytical solutions, 
however, are the same (see for example the right panel of Figure 1 in \citet{ert09}). 
In reality, in order that all inflowing mass accretes, 
the boundary condition $\Sigma (r_{\mathrm{in}},t)=0$ must be satisfied. 
The exact self-similar solution overestimates 
the surface mass density near the inner region such that this extra matter 
at the inner region compensates the reduction in the mass flow rate due to reduced density gradient.
The right panel of Figure~\ref{disc_uni}  stands for the full propeller solution. 
In this case we see that the exact solution given in Equation~(\ref{solut2}) does 
very fine except that it extends to the origin though real discs have finite radius. 
The outer radius of the disc, in order to take up the angular momentum imparted from 
the inner boundary, moves out more rapidly in this case.

The radial profile of the full propeller solution given in Equation~(\ref{solut2}), $\sigma \propto R^{-(p+1/2)/(q+1)}$, is steeper than the radial profile of the full accretor solution given in Equation~(\ref{solut1}), $\sigma \propto R^{-p/(q+1)}$ \citep{spr93}. This leads to enhanced accretion in transitions from the propeller to the accreting regimes. This then will lead to the inwards movement of the inner disc radius $r_{\rm in}$ leading to cyclic behavior in systems at the brink of transitions between accretor and propeller regimes \citep{spr93,dang12}.

\section{RESULTS: A SOLUTION FOR PARTIAL ACCRETION}
In this section we demonstrate that the approximate solution given in Equation~(\ref{approximate}) 
describes discs with partial accretion accurately.

\subsection{Numerical Method}

In solving the diffusion equation (\ref{diff1}) numerically we first transform to the ``specific angular momentum coordinate'' $x\equiv r^{1/2}$ and use $S\equiv x\Sigma$ \citep{bat81}. The equation becomes
\begin{equation}
\frac{\partial S }{\partial t }=\frac{1}{x^2}\frac{\partial^2 }{\partial
x^2} ( \nu S ) . 
\label{diff3}
\end{equation}
where $\nu = C x^{2p-q}S^q$. With this notation Equation~(\ref{acc_rate}) becomes
\begin{equation}
\dot{M}=3\pi \frac{\partial }{\partial x} ( \nu S ) . 
\label{acc_rate3}
\end{equation}
These equations are discretized and the new value of $S$ at each grid point is found by
\begin{equation}
S^{\rm new}_i =S_i + \frac{\Delta t}{(x_i \Delta x)^2}\left[ (\nu S)_{i+1}+(\nu S)_{i-1}-2(\nu S)_i  \right].
\label{disc_s}
\end{equation}
The  discretization of Eqn.\ (\ref{acc_rate3})) gives accretion rate at each grid point as
\begin{equation}
\dot{M}_i =3\pi \frac{(\nu S)_{i+1}-(\nu S)_i}{\Delta x}.
\label{disc_m}
\end{equation}
We have used $i_{\max}=2000$ grids $x_i = x_0 + i\Delta x$ where  $x_0=r_{0}^{1/2}$. We have
taken $r_{0}=10^8$ cm and $r_{i_{\max}}=10^{13}$ cm.

\begin{figure*}
\includegraphics[width=0.45\textwidth]{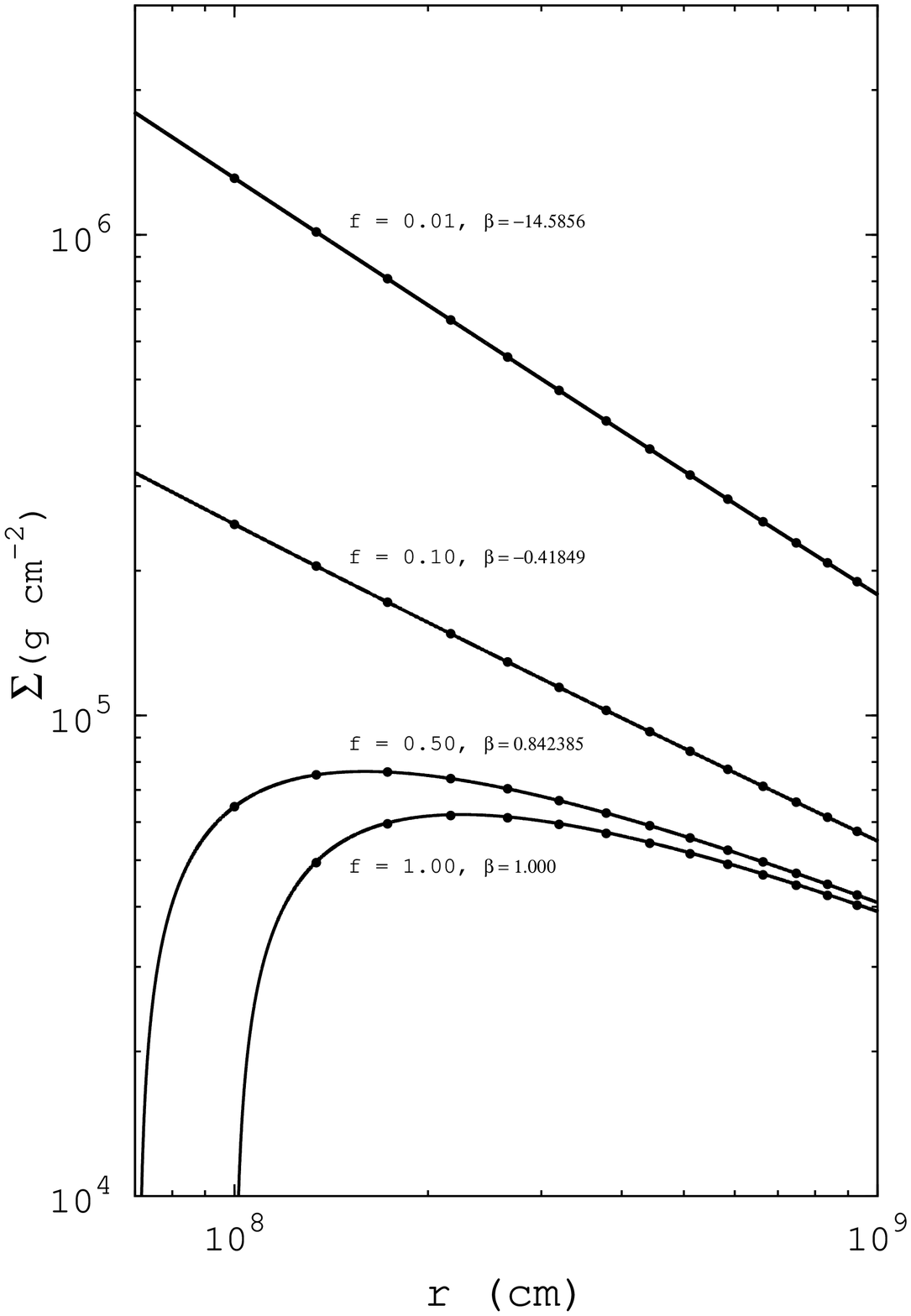}
\includegraphics[width=0.45\textwidth]{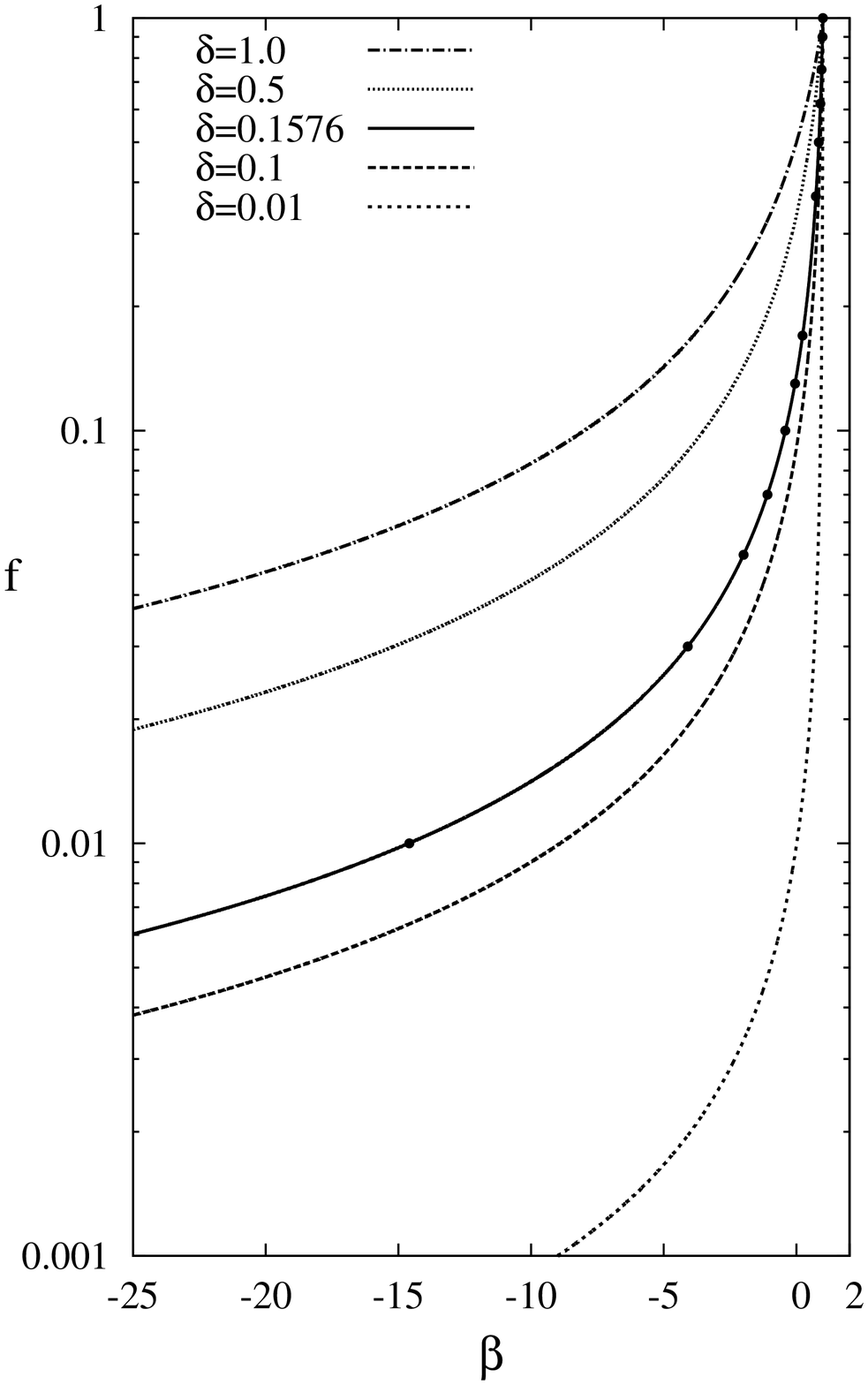}
\caption{The left panel shows the surface mass density profile for discs with partial accretion. The cases for $f=0.01$, 0.1, 0.5 and 1 are shown. 
The right panel shows the relation between the fraction of mass flux, $f$, that traverses the inner boundary of the disc and the dimensionless torque at the inner boundary, $\beta$ as given in Equation~(\ref{width}). The data points are the $\beta$ values obtained by fitting the surface mass density given in Equation~(\ref{approximate}) to the numerical solution for a value of $f$.
In the Figure we have not applied any fitting procedure, but used $\delta \equiv \Delta x/x_{\rm in}=0.1576$ from our simulation parameters.}
\label{fig_sig}
\end{figure*}

\subsection{Inner Boundary Condition for Partial Accretion}

The mass flux at the inner boundary of the disc ($i=0$) is
\begin{equation}
\dot{M}_{\ast} =3\pi \frac{(\nu S)_1-(\nu S)_0}{\Delta x}
\label{disc_in}
\end{equation}
Accordingly, the maximum mass flux is obtained for $(\nu S)_0=0$, the full accretor boundary condition, as
\begin{equation}
\dot{M}_{\max} =3\pi \frac{(\nu S)_1}{\Delta x}.
\label{mdot_max}
\end{equation}
We have used the boundary condition $(\nu S)_0=0$ for the full accretion regime 
(see left panel in Figure~\ref{disc_uni}). 

According to Equation (\ref{disc_in}) again, $\dot{M}_{\ast}=0$ if  $(\nu S)_0=(\nu S)_1$ 
and we have employed this boundary condition 
for the full propeller regime (see right panel in Figure~\ref{disc_uni}).
If a fraction $f$ of $\dot{M}_{\max}$ accretes
\begin{equation}
f \dot{M}_{\max} =3\pi \frac{(\nu S)_1-(\nu S)_0}{\Delta x}
\end{equation}
is satisfied which implies 
\begin{equation}
(\nu S)_0=(1 - f)(\nu S)_1.
\label{partial_BC}
\end{equation}
In the following, we employ this boundary condition for simulating the evolution of discs in 
which a fraction $f$ of the inflowing mass flux traverses the inner boundary
and the rest $1-f$ remains in the disc. 
Note that $f=1$ and $f=0$ cases correspond to the well know full accretor and 
full propeller boundary conditions, respectively.

With $x=r^{1/2}$ and $S=\Sigma x$ Equation~(\ref{Jdot3}) can be written as 
\begin{equation}
\nu S = \frac{\dot{M}}{3\pi}(x-\beta x_{\rm in})
\end{equation}
Finding $(\nu S)_0$ and $(\nu S)_1$ from this equation and plugging into Equation (\ref{partial_BC}) gives
\begin{equation}
\delta  = \frac{1-\beta}{1-f} f
\label{width}
\end{equation}
where 
\begin{equation}
\delta \equiv \frac{\Delta x}{x_{\rm in}} = \frac{\Delta r}{2 r_{\rm in}}
\label{delta}
\end{equation}
which is
the relative width of grids in the specific angular momentum space.
For our simulation parameters $\delta =0.1576$.

We have performed simulations of disc evolution for a range of $f$ values from 0.01 to 1 keeping the value of $f$ constant throughout the simultation. For each value of $f$ we have seen that the self-similar evolution of the disc is well described by the approximate solution given in Equation~(\ref{approximate}), 
for a certain value of $\beta$ found by solving $\beta$ from Equation~(\ref{width}):
\begin{equation}
\beta = 1 - \frac{1-f}{f}\delta.
\label{nf}
\end{equation}
The left panel of Figure~\ref{fig_sig} shows the surface density profile in the inner region of the disc 
at $t=0.1$ years for $f=0.01$, 0.10, 0.50 and 1.00. The corresponding $\beta$ values are also written in the Figure. The right panel of Figure~\ref{fig_sig} shows the relation given in Equation~(\ref{width}) together with data points from our numerical simulations. The Figure also shows results for other values of $\delta$ from 0.01 to 1.

The relative width given in Equation (\ref{width}) is determined by the number of grids we have chosen. 
It thus appears as a numerical quantity that would vanish if we could do numerical computing with 
infinite number of grids. At the inner boundary of the disc 
the matter rotating with keplerian angular velocity has to come to corotation with the 
star over some region of narrow width, the boundary layer.
In our numerical solution of the diffusion equation (\ref{diff1}) in which the Keplerian rotation is built in, 
non-keplerian rotation in the boundary layer can not be modeled and it is implicitly  assumed that
this transition takes place at the innermost grid and is unresolved. For real discs which do not have grids,
$\delta$ has the physical meaning as
the relative width of the region over which matter in the disc is brought into corotation with the star.
Equation (\ref{width}) then tells that the relative width of this transition region is determined by 
the fraction of matter that could traverse it and the dimensionless torque. 
The relative width is determined by the stresses that bring the matter in the disc to co-rotation with the star. It is not necessarily
the viscous torques.  In the case of magnetized stars truncating the disc beyond the stellar surface this could be the magnetic stress applied 
by the rotating magnetosphere of the star. 

\subsection{Partial Accretion Stage Away From Equilibrium}

In the previous subsection the value of $f$ was kept constant throughout the simulation.
It was found that the accretion rate leaving the disc $\dot{M}_{\ast}$ 
follows precisely the value given in Equation~(\ref{Mdot_d}) calculated from the full accretor solution
and not less of it. 
This is because when a fraction $f$ of the inflowing mass accretes and the rest $1-f$ remains in the disc the
surface density $\Sigma$ and subsequently the accretion rate increases. Within a viscous time-scale in 
the disc $\dot{M}_{\max}$ increases so that $f \dot{M}_{\max}=\dot{M}_{\rm d}$ and not something less.
Self-similar approximate solution given in Equation~(\ref{approximate}) describes this asymptotic stage. 

\begin{figure}
\includegraphics[width=1.0\textwidth]{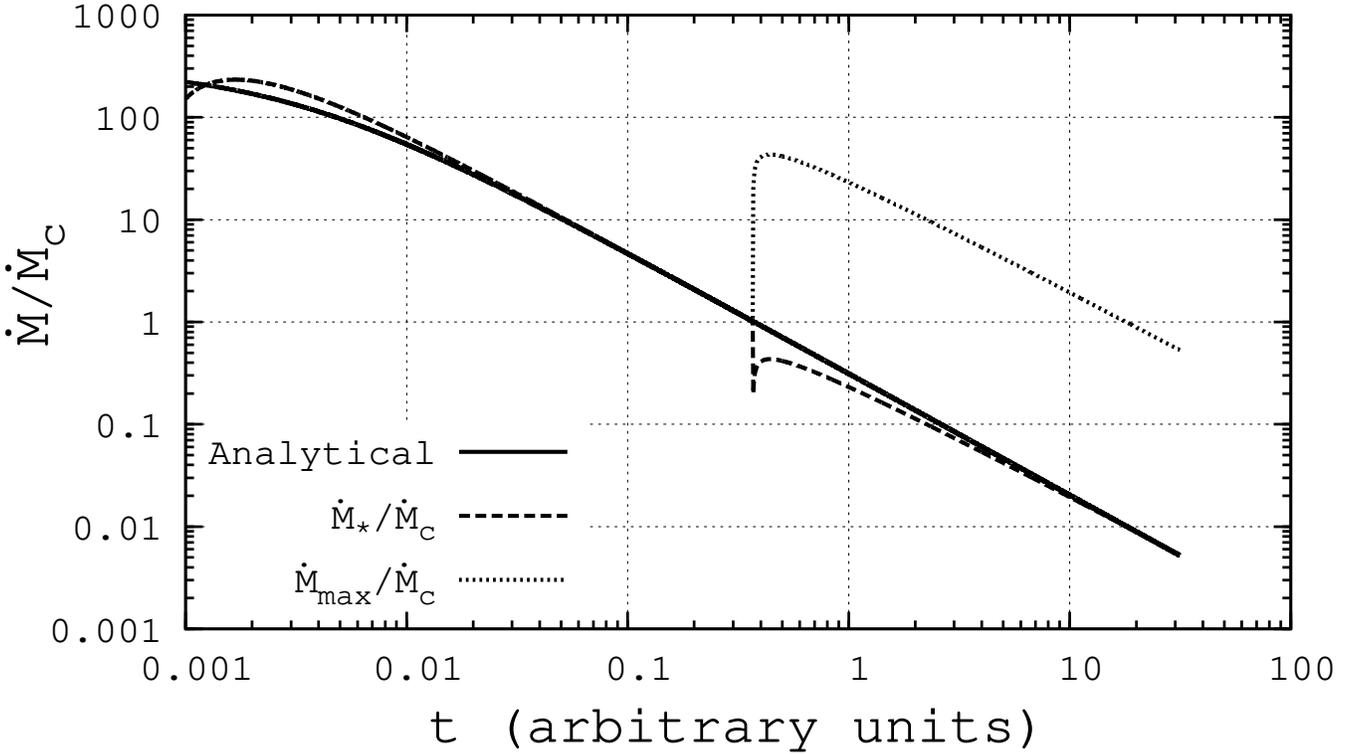}
\caption{Mass accretion rate in units of critical mass flow rate $\dot{M}_{\rm c}$ 
at which partial accretion starts. The solid line shows the analytical expression of mass accretion 
rate given in Equation~(\ref{Mdot_d}). Initially $\dot{M}>\dot{M}_{\rm c}$ and so all mass accretes ($f=1$).
The dashed lines showing $\dot{M}_{\max}$ (see Eqn.(\ref{mdot_max})) and dotted lines showing $\dot{M}_{\ast}$ in units of $\dot{M}_{\rm c}$ 
coincide at this stage and they both catch up with the analytical expression soon after the initial conditions are forgotten.
When the mass inflow rate drops below $\dot{M}_{\rm c}$ only a fraction $f=0.01$ is allowed to accrete onto the central object
using the boundary condition given in Equation~(\ref{partial_BC}). Although $\dot{M}_{\ast}$ initially drops abruptly as a result
of this boundary condition, it catches up with the analytical value given in Equation~(\ref{Mdot_d}) in a viscous time-scale. 
This is because the density at the inner disc increases in response as 99\% of the inflowing mass is retained in the disk i.e.\ $\dot{M}_{\max}$ tends to increase to $\dot{M}_{\rm d}/f$ so that its multiplication with $f$ still gives $\dot{M}_{\rm d}$.
}
\label{fig_Mdot_in}
\end{figure}

In order to see the behavior of the system when the boundary condition changes we should change the boundary condition more
rapidly than the viscous time-scale in the disc.  With this motivation we have changed $f$ from 1 to 0.01 abruptly at a critical 
mass flow rate $\dot{M}_{\rm c}$ at which centrifugal barrier sets in. 
In real systems, of course, we expect that $f$ will change more smoothly with the fastness parameter \citep[see e.g.\ ][]{eks10}. The result of this simulation is shown in Figure~\ref{fig_Mdot_in}. 
We observe that with the start of the partial accretion regime the accretion rate onto the star deviates from its trend given by Equation~(\ref{Mdot_d}). This is only for a brief epoch because
the rest of the matter retained in the disc increases $\dot{M}_{\max}$ to $\dot{M}_{\rm d}/f$ so that 
$\dot{M}_{\ast}=f \dot{M}_{\max}$ asymptotes to the value given by Equation~(\ref{Mdot_d}). The transient stage following the immediate
aftermath of the transition to partial accretion is not modeled by the approximate self-similar solution given in 
Equation (\ref{approximate}) as self-similar solutions describe the asymptotic stage when the system reaches 
quasi-equilibrium.

\section{DISCUSSION} 

By using a recently suggested prescription we have constructed an approximate self-similar 
solution, given in Equation (\ref{unified}), describing the evolution of a partially accreting disc. 
This motivated another approximate solution given in Equation (\ref{approximate}) which has a constant finite inner radius.
We have numerically shown that this solution describes the asymptotic stage achieved 
after a few viscous time-scales just like the exact solutions but describes the continuous range of inner boundary conditions between full accretion and full propeller. The approximate solution describes the inner region of the disk even better than the exact solution in the full accretion stage ($f=1$).

The partial accretion might lead to a rapid decline stage \citep{ibr09} during an outburst of a neutron star 
soft X-ray transient when the accretion rate drops below a critical value at which the inner radius moves beyond the corotation radius and the centrifugal barrier allows only a fraction of the inflowing matter from the off midplane of the disc \citep{eks10}. Our results imply that partial accretion stage can change the accretion luminosity only if the decay timescale is short so that the inner boundary condition changes abruptly. Otherwise the mass accumulation at the inner boundary increases the accretion rate onto the central object to the same value it would have if there was no partial accretion within the viscous time-scale. As the viscous time scale in the disc increases with the size of the disc it leads to the expectation that partial accretion can lead to a rapid decline stage in systems
allowing small size discs i.e.\ transient systems with short orbital periods like accreting millisecond pulsars.

The self-similar solutions considered here are more appropriate for free discs e.g.\ supernova debris discs around young neutron stars \citep{mic81,alp01,cha00} rather than the tidally truncated discs in binary systems. A debris disc was discovered by \citet{wan06} around an anomalous X-ray pulsar (AXP;  see \citet{mer08} for a review). Central compact objects  \citep[see][for a review]{deL08} like Cas A might also harbor fallback discs \citep{alp01}. \citet{tag03} mentioned that fallback disc models of AXPs should incorporate accretion and propeller mechanisms working simultaneously. Partial accretion was proposed by \citet{eks03} to explain the present day luminosities of AXPs if they have descended from 10 ms spin periods by interacting with fallback discs.
Our results imply for these systems that even if some of these systems are in a partial accretion regime, their long term luminosity trend 
is determined by the mass flow rate derived from the full accretion solution of \citet{pri74a}. Mass flow rate variations due to disc instabilities would result in modifications of this accretion rate if these changes occur sufficiently rapid.


\begin{thebibliography}{}

\bibitem[\protect\citeauthoryear{{Alpar}}{{Alpar}}{2001}]{alp01}
{Alpar} M.~A.,  2001, \apj, 554, 1245

\bibitem[\protect\citeauthoryear{{Barenblatt}}{{Barenblatt}}{1996}]{bar96}
{Barenblatt} G.~I.,  1996, {Scaling, Self-similarity, and Intermediate
  Asymptotics}

\bibitem[\protect\citeauthoryear{{Bath} \& {Pringle}}{{Bath} \&
  {Pringle}}{1981}]{bat81}
{Bath} G.~T.,  {Pringle} J.~E.,  1981, \mnras, 194, 967

\bibitem[\protect\citeauthoryear{{Cannizzo}, {Lee} \& {Goodman}}{{Cannizzo}
  et~al.}{1990}]{can90}
{Cannizzo} J.~K.,  {Lee} H.~M.,    {Goodman} J.,  1990, \apj, 351, 38

\bibitem[\protect\citeauthoryear{{Chan}, {Psaltis} \& {{\"O}zel}}{{Chan}
  et~al.}{2005}]{cha05}
{Chan} C.-k.,  {Psaltis} D.,    {{\"O}zel} F.,  2005, \apj, 628, 353

\bibitem[\protect\citeauthoryear{{Chatterjee}, {Hernquist} \&
  {Narayan}}{{Chatterjee} et~al.}{2000}]{cha00}
{Chatterjee} P.,  {Hernquist} L.,    {Narayan} R.,  2000, \apj, 534, 373

\bibitem[\protect\citeauthoryear{{D'Angelo} \& {Spruit}}{{D'Angelo} \&
  {Spruit}}{2011}]{dang11}
{D'Angelo} C.~R.,  {Spruit} H.~C.,  2011, \mnras, 416, 893

\bibitem[\protect\citeauthoryear{{D'Angelo} \& {Spruit}}{{D'Angelo} \&
  {Spruit}}{2012}]{dang12}
{D'Angelo} C.~R.,  {Spruit} H.~C.,  2012, \mnras, 420, 416

\bibitem[\protect\citeauthoryear{{de Luca}}{{de Luca}}{2008}]{deL08}
{de Luca} A.,  2008, in {C.~Bassa, Z.~Wang, A.~Cumming, \& V.~M.~Kaspi} ed., 40
  Years of Pulsars: Millisecond Pulsars, Magnetars and More Vol.~983 of
  American Institute of Physics Conference Series, {Central Compact Objects in
  Supernova Remnants}.
pp 311--319

\bibitem[\protect\citeauthoryear{{Ek\c{s}i}}{{Ek\c{s}i}}{2009}]{eks09}
{Ek\c{s}i} K.~Y.,  2009, ArXiv e-prints: 0908.3337

\bibitem[\protect\citeauthoryear{{Ek\c{s}i} \& {Alpar}}{{Ek\c{s}i} \&
  {Alpar}}{2003}]{eks03}
{Ek\c{s}i} K.~Y.,  {Alpar} M.~A.,  2003, \apj, 599, 450

\bibitem[\protect\citeauthoryear{{Ek\c{s}i} \& {Kutlu}}{{Ek\c{s}i} \&
  {Kutlu}}{2010}]{eks10}
{Ek\c{s}i} K.~Y.,  {Kutlu} E.,  2010, ArXiv e-prints: 1010.1528

\bibitem[\protect\citeauthoryear{{Ertan}, {Ek{\c s}i}, {Erkut} \&
  {Alpar}}{{Ertan} et~al.}{2009}]{ert09}
{Ertan} {\"U}.,  {Ek{\c s}i} K.~Y.,  {Erkut} M.~H.,    {Alpar} M.~A.,  2009,
  \apj, 702, 1309

\bibitem[\protect\citeauthoryear{{Frank}, {King} \& {Raine}}{{Frank}
  et~al.}{2002}]{fra02}
{Frank} J.,  {King} A.,    {Raine} D.~J.,  2002, {Accretion Power in
  Astrophysics: Third Edition}

\bibitem[\protect\citeauthoryear{{Ibragimov} \& {Poutanen}}{{Ibragimov} \&
  {Poutanen}}{2009}]{ibr09}
{Ibragimov} A.,  {Poutanen} J.,  2009, \mnras, 400, 492

\bibitem[\protect\citeauthoryear{{Illarionov} \& {Sunyaev}}{{Illarionov} \&
  {Sunyaev}}{1975}]{ill75}
{Illarionov} A.~F.,  {Sunyaev} R.~A.,  1975, \aap, 39, 185

\bibitem[\protect\citeauthoryear{{Lynden-Bell} \& {Pringle}}{{Lynden-Bell} \&
  {Pringle}}{1974}]{pri74b}
{Lynden-Bell} D.,  {Pringle} J.~E.,  1974, \mnras, 168, 603

\bibitem[\protect\citeauthoryear{{Menou}, {Esin}, {Narayan}, {Garcia}, {Lasota}
  \& {McClintock}}{{Menou} et~al.}{1999}]{men99}
{Menou} K.,  {Esin} A.~A.,  {Narayan} R.,  {Garcia} M.~R.,  {Lasota} J.-P.,
  {McClintock} J.~E.,  1999, \apj, 520, 276

\bibitem[\protect\citeauthoryear{{Mereghetti}}{{Mereghetti}}{2008}]{mer08}
{Mereghetti} S.,  2008, \aapr, 15, 225

\bibitem[\protect\citeauthoryear{{Michel} \& {Dessler}}{{Michel} \&
  {Dessler}}{1981}]{mic81}
{Michel} F.~C.,  {Dessler} A.~J.,  1981, \apj, 251, 654

\bibitem[\protect\citeauthoryear{{Mineshige}}{{Mineshige}}{1991}]{min91}
{Mineshige} S.,  1991, \mnras, 250, 253

\bibitem[\protect\citeauthoryear{{Mineshige}, {Nomoto} \&
  {Shigeyama}}{{Mineshige} et~al.}{1993}]{min93}
{Mineshige} S.,  {Nomoto} K.,    {Shigeyama} T.,  1993, \aap, 267, 95

\bibitem[\protect\citeauthoryear{{Perna}, {Hernquist} \& {Narayan}}{{Perna}
  et~al.}{2000}]{per00}
{Perna} R.,  {Hernquist} L.,    {Narayan} R.,  2000, \apj, 541, 344

\bibitem[\protect\citeauthoryear{{Pringle}}{{Pringle}}{1974}]{pri74a}
{Pringle} J.~E.,  1974, PhD thesis, , Univ.~Cambridge, (1974)

\bibitem[\protect\citeauthoryear{{Pringle}}{{Pringle}}{1981}]{pri81}
{Pringle} J.~E.,  1981, \araa, 19, 137

\bibitem[\protect\citeauthoryear{{Pringle}}{{Pringle}}{1991}]{pri91}
{Pringle} J.~E.,  1991, \mnras, 248, 754

\bibitem[\protect\citeauthoryear{{Pringle} \& {Rees}}{{Pringle} \&
  {Rees}}{1972}]{pri72}
{Pringle} J.~E.,  {Rees} M.~J.,  1972, \aap, 21, 1

\bibitem[\protect\citeauthoryear{{Romanova}, {Ustyugova}, {Koldoba} \&
  {Lovelace}}{{Romanova} et~al.}{2004}]{rom04}
{Romanova} M.~M.,  {Ustyugova} G.~V.,  {Koldoba} A.~V.,    {Lovelace} R.~V.~E.,
   2004, \apjl, 616, L151

\bibitem[\protect\citeauthoryear{{Shakura} \& {Sunyaev}}{{Shakura} \&
  {Sunyaev}}{1973}]{sha73}
{Shakura} N.~I.,  {Sunyaev} R.~A.,  1973, \aap, 24, 337

\bibitem[\protect\citeauthoryear{{Spruit} \& {Taam}}{{Spruit} \&
  {Taam}}{1993}]{spr93}
{Spruit} H.~C.,  {Taam} R.~E.,  1993, \apj, 402, 593

\bibitem[\protect\citeauthoryear{{Spruit} \& {Taam}}{{Spruit} \&
  {Taam}}{2001}]{spr01}
{Spruit} H.~C.,  {Taam} R.~E.,  2001, \apj, 548, 900

\bibitem[\protect\citeauthoryear{{Sunyaev} \& {Shakura}}{{Sunyaev} \&
  {Shakura}}{1977}]{sun77}
{Sunyaev} R.~A.,  {Shakura} N.~I.,  1977, Pis ma Astronomicheskii Zhurnal, 3,
  262

\bibitem[\protect\citeauthoryear{{Tagieva}, {Yazgan} \& {Ankay}}{{Tagieva}
  et~al.}{2003}]{tag03}
{Tagieva} S.~O.,  {Yazgan} E.,    {Ankay} A.,  2003, International Journal of
  Modern Physics D, 12, 825

\bibitem[\protect\citeauthoryear{{Tanaka}}{{Tanaka}}{2011}]{tan11}
{Tanaka} T.,  2011, \mnras, 410, 1007

\bibitem[\protect\citeauthoryear{{Ustyugova}, {Koldoba}, {Romanova} \&
  {Lovelace}}{{Ustyugova} et~al.}{2006}]{ust06}
{Ustyugova} G.~V.,  {Koldoba} A.~V.,  {Romanova} M.~M.,    {Lovelace} R.~V.~E.,
   2006, \apj, 646, 304

\bibitem[\protect\citeauthoryear{{Wang}, {Chakrabarty} \& {Kaplan}}{{Wang}
  et~al.}{2006}]{wan06}
{Wang} Z.,  {Chakrabarty} D.,    {Kaplan} D.~L.,  2006, \nat, 440, 772

\bibitem[\protect\citeauthoryear{{Zel'Dovich} \& {Raizer}}{{Zel'Dovich} \&
  {Raizer}}{1967}]{zel67}
{Zel'Dovich} Y.~B.,  {Raizer} Y.~P.,  1967, {Physics of shock waves and
  high-temperature hydrodynamic phenomena}

\end{thebibliography}

\label{lastpage}

\end{document}